\documentclass[10pt,pre,twocolumn,byrevtex,bibnotes,showpacs]{revtex4}
\usepackage{amsfonts}
\usepackage{graphics}
\usepackage{amsthm}

\theoremstyle{definition}
\newtheorem{theorem}{Theorem}
\newtheorem{definition}[theorem]{Definition}
\begin{document}

\title{Nonconcave entropies in multifractals and the thermodynamic formalism}
\author{Hugo Touchette}
\email{htouchet@alum.mit.edu}
\affiliation{School of Mathematical Sciences, Queen Mary,
University of London, London E1 4NS, UK}

\author{Christian Beck}
\email{c.beck@qmul.ac.uk}
\affiliation{School of Mathematical Sciences, Queen Mary,
University of London, London E1 4NS, UK}
\date{\today}

\begin{abstract}
We discuss a subtlety involved in the calculation of multifractal spectra
when these are expressed as Legendre-Fenchel transforms of functions
analogous to free energy functions. We show that the Legendre-Fenchel
transform of a free energy function yields the correct multifractal spectrum
only when the latter is wholly concave. If the spectrum has no definite
concavity, then the transform yields the concave envelope of the spectrum
rather than the spectrum itself. Some mathematical and physical examples are
given to illustrate this result, which lies at the root of the nonequivalence
of the microcanonical and canonical ensembles. On a more positive note, we also
show that the impossibility of expressing nonconcave multifractal
spectra through Legendre-Fenchel transforms of free energies can be
circumvented with the help of a generalized free energy function, which
relates to a recently introduced generalized canonical ensemble. Analogies
with the calculation of rate functions in large deviation theory are finally
discussed.
\end{abstract}
\maketitle

\section{Introduction}

Invariant measures generated by nonlinear and complex dynamical systems
often show striking scaling and self-similar features that are reminiscent
of fractals. However, contrary to ordinary fractals, whose geometric
structure is characterized by a single number (the fractal or Hausdorff
dimension \cite{falconer1990}), the scaling and self-similar properties of
measures are usually not captured by a single dimension, say $\alpha $, but
by an infinite set of fractal or singularity dimensions that defines the
so-called \textit{spectrum of singularities} $f(\alpha )$, also known as the 
\textit{multifractal spectrum} \cite{beck1993,mandelbrot1999}. The word
``multifractal'' has been coined \cite{frisch1985} in this context precisely
to suggest that a measure having multiscaling properties can be pictured
abstractly as a superposition of many ``pure'' fractals, each having a
dimension $\alpha $ and a corresponding ``weight'' $f(\alpha )$ in the
superposition.

To be more specific, consider a measure $\mu $ defined on a $d$-dimensional
space $X$. Generalizing the approach followed in fractal geometry, we
proceed to partition or ``coarse-grain'' the space $X$ in small boxes of
equal size $\varepsilon $ and volume $\varepsilon ^d$. The measure contained
in each box is 
\begin{equation}
p_{\varepsilon ,i}=\int_{i^{\text{th}}\text{ box}}d\mu (x),
\end{equation}
and from this quantity, a local fractal dimension $\alpha _i$, also called a
crowding index, is defined by using the fact that $p_{\varepsilon ,i}$ is
expected to scale as $p_{\varepsilon ,i}\sim \varepsilon ^{\alpha _i}$ in
the limit where the boxes' size $\varepsilon $ goes to zero. Now, to account
for the fact that $\alpha _i$ is not constant over the partition but varies
in general from one box to another, we count the number $n_\varepsilon
(\alpha )$ of boxes in the partition whose local dimension is equal to
$\alpha $. From $n_\varepsilon (\alpha )$, the multifractal spectrum
$f(\alpha )$ is then simply defined through another scaling relationship,
namely $n_\varepsilon (\alpha )\sim \varepsilon ^{-f(\alpha )}$ as
$\varepsilon \rightarrow 0$.

The multifractal spectrum $f(\alpha )$ is not a quantity which is easily
calculated analytically or numerically, since it requires the enumeration of
all the boxes in the partition of $X$ having a crowding index $\alpha $
lying in some interval $[\alpha ,\alpha +\Delta \alpha ]$. A more manageable
quantity which can be related to $f(\alpha )$ is the so-called \textit{free
energy function} $\tau (q)$ defined by the scaling relationship
$Z_\varepsilon (q)\sim \varepsilon ^{\tau (q)}$, $\varepsilon \rightarrow 0$,
where 
\begin{equation}
Z_\varepsilon (q)=\sum_ip_{\varepsilon ,i}^q\sim \sum_i\varepsilon
^{-q\alpha _i}
\end{equation}
is the \textit{partition function} associated with the partition $X$ of $\mu$ 
(the sum above runs over all the boxes of the partition with 
$p_{\varepsilon ,i}\neq 0$ since $q$ can be negative). The calculation of
$\tau (q)$ parallels the calculation of free energies in statistical
mechanics in that, if $f(\alpha )$ is known, then $\tau (q)$ can be
calculated as the \textit{Legendre-Fenchel} (LF) \textit{transform} of
$f(\alpha )$ \cite{beck1993}; in symbols, 
\begin{equation}
\tau (q)=\inf_{\alpha \in \Bbb{R}}\{q\alpha -f(\alpha )\}.  \label{LF1}
\end{equation}

\begin{figure*}[t]
\centering
\includegraphics{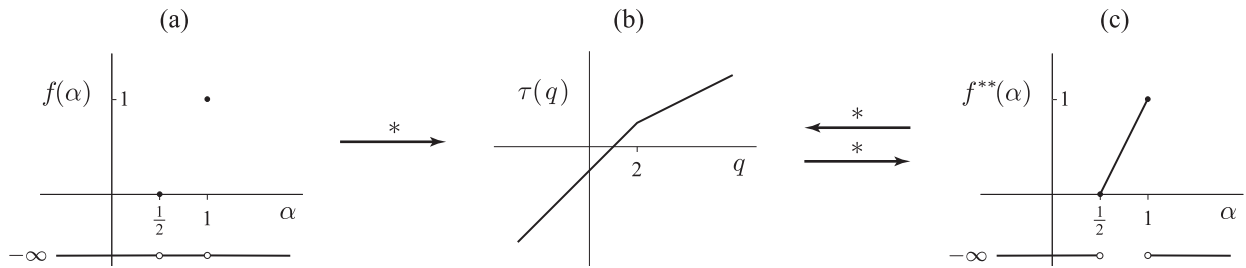}
\caption{(a) Multifractal spectrum $f(\protect\alpha )$ for the invariant
density of the Ulam map. (b) Corresponding free energy function $\protect\tau(q)$.
(c) Legendre-Fenchel transform of $\protect\tau (q)$.}
\label{exfig1}
\end{figure*}

The result that we shall study in this paper is the inverse result, namely
that if $\tau (q)$ is known, then $f(\alpha )$ can be calculated from $\tau
(q)$ by taking the LF transform of the latter function; in symbols, 
\begin{equation}
f(\alpha )=\inf_{q\in \Bbb{R}}\{q\alpha -\tau (q)\}.  \label{LF2}
\end{equation}
This result first appeared in Refs.~\cite{frisch1985,halsey1986}, and has
been used extensively since then to calculate the multifractal spectrum of
many phenomena, including turbulence 
\cite{benzi1984,benzi1991,benzi1993,stolovitzky1993,schertzer1985,schertzer1997},
geophysical processes, such as cloud formation and rain precipitations
\cite{schertzer1987,schertzer1989,schertzer1991}, and fluctuations in financial
time series \cite{schmitt1999,mandelbrot1999}, among many others
\cite{paladin1987}. Unfortunately, there is one aspect of Eq.(\ref{LF2}) which is
often overlooked when deriving it and applying it, namely that \textit{it
can only produce concave multifractal spectra}, since LF transforms can only
yield concave functions. This basic property of LF transforms does not
affect, as such, the calculation of $\tau (q)$ from $f(\alpha )$ because it
can be proved that $\tau (q)$ is an always concave function of $q$. For
calculating the multifractal spectrum, however, there is a problem because 
$f(\alpha )$ need not be concave, which means that $f(\alpha )$ cannot always
be calculated as the LF transform of $\tau (q)$.

Our goal here is to illustrate these observations with a number of basic
examples, and to state the precise conditions, based on convex analysis,
that ensure that $f(\alpha )$ can be calculated as the LF transform of
$\tau(q)$. These conditions will be discussed in the context of four
physically-relevant multifractal models: one related to turbulence, another
related to diffusion-limited aggregates, and two others related to chaotic
systems. In an attempt to offer a workable solution to the problem of
calculating nonconcave multifractal spectra, we shall also study a recently introduced
generalized canonical ensemble to show that nonconcave spectra can be
obtained from a modified version of the LF transform. This part will
actually provide an explicit calculation of a nonconcave spectrum $f(\alpha )$
which uses this modified LF transform. We shall comment finally,
in the concluding section of the paper, on analogies between nonconcave
multifractal spectra, nonconcave entropies in statistical mechanics, and
large deviation theory.

\section{Two simple examples}
\label{sExamples}

We begin by considering two explicit examples of measures whose multifractal
spectra are not given by LF transforms of their free energy functions. The
first example was previously discussed in Ref.~\cite{ott1984} (see also 
Ref.~\cite{beck1993}), and will serve here as a starting point to our discussion
of the validity of the LF\ transform of (\ref{LF2}). The measure or, rather,
the density in this case that we consider is given by 
\begin{equation}
\rho (x)=\frac 1{\pi \sqrt{1-x^2}},
\end{equation}
where $x\in [-1,1]$. This density arises as the invariant density of the
Ulam map and the Tchebyscheff maps. Applying a partition of size
$\varepsilon $ on the interval $[-1,1]$, it can be seen that the two boxes of
the partition located near the boundary points $x=\pm 1$ have measure 
$p_{\varepsilon ,i}=\varepsilon \rho (x)\sim \varepsilon ^{1/2}$, so that
$f(\alpha )=0$ at $\alpha =1/2$. All the other boxes have measure 
$p_{\varepsilon ,i}\sim \varepsilon $, so that $f(\alpha )=1$ at $\alpha =1$,
as there are approximately $n_\varepsilon (\alpha )\sim \varepsilon ^{-1}$
of these boxes. Combining the two results, and setting $n_\varepsilon
(\alpha )=0$ for $\alpha \notin \{1/2,1\}$, we obtain 
\begin{equation}
f(\alpha )=\left\{ 
\begin{array}{lll}
0 &  & \alpha =1/2 \\ 
1 &  & \alpha =1 \\ 
-\infty  &  & \text{otherwise.}
\end{array}
\right.   \label{initf1}
\end{equation}
This spectrum is shown in Fig.~\ref{exfig1}(a). 

At this point, we go on to prove that $f(\alpha )$ cannot be expressed as the
LF transform of $\tau (q)$ by direct calculation. 
Starting from the asymptotic ($\varepsilon
\rightarrow 0$) expression of the partition function 
\begin{equation}
Z_\varepsilon (q)=\sum_ip_{\varepsilon ,i}^q\sim \varepsilon
^{q/2}+\varepsilon ^{-1}\varepsilon ^q,
\end{equation}
we first find 
\begin{equation}
\tau (q)=\min \{q-1,q/2\}=\left\{ 
\begin{array}{lll}
q/2 &  & q>2 \\ 
q-1 &  & q\leq 2.
\end{array}
\right. 
\end{equation}
Then keeping track of the two separate regions $q>2$ and $q\leq 2$, we find 
\begin{eqnarray}
\inf_{q\in \Bbb{R}}\{q\alpha -\tau (q)\} &=&\inf_{q\in \Bbb{R}}\left\{ 
\begin{array}{lll}
q(\alpha -\frac 12) &  & q>2 \\ 
q(\alpha -1)+1 &  & q\leq 2
\end{array}
\right\}   \nonumber \\
&=&\left\{ 
\begin{array}{lll}
2\alpha -1 &  & \alpha \in [1/2,1] \\ 
-\infty  &  & \text{otherwise.}
\end{array}
\right.   \label{finf1}
\end{eqnarray}
Comparing this result with Eq.(\ref{initf1}), we see that $f(\alpha )$
corresponds to the LF\ transform of $\tau (q)$ for $\alpha \notin (1/2,1)$ only; see Figs.~\ref{exfig1}(a) and \ref{exfig1}(c). For
$\alpha\in(1/2,1)$, the LF transform of $\tau(q)$ is finite, while the true 
spectrum $f(\alpha)$ is formally equal to $-\infty$, as there is no box in the partition
of $\rho(x)$ with local exponent in the range $(1/2,1)$.

This example can be generalized to illustrate another problem when trying to 
obtain $f(\alpha )$ from $\tau (q)$. Consider a dynamical
system in $d$-dimensions whose invariant density is everywhere finite, so
that $p_{i,\varepsilon }\sim \varepsilon ^d$,\ except at a finite number $k$
of singular points where $p_{i,\varepsilon }\sim \varepsilon ^{\alpha_i}=
\varepsilon ^{d\xi _i}$ with $\xi _1,\xi _2,\ldots ,\xi _k<1$. The
partition function for this density is 
\begin{equation}
Z_\varepsilon (q)=\sum_ip_i^q\sim \varepsilon ^{-d}\varepsilon
^{dq}+\varepsilon ^{d\xi _1q}+\varepsilon ^{d\xi _2q}+\cdots +\varepsilon
^{d\xi _kq},
\end{equation}
so that 
\begin{equation}
\tau (q)=\min \{(q-1)d,d\xi _1q,d\xi _2q,\ldots ,d\xi _kq\}.
\end{equation}
The minimum can be calculated explicitly and yields 
\begin{equation}
\tau (q)=\left\{ 
\begin{array}{lll}
d(q-1) &  & q\leq (1-\xi ^{*})^{-1} \\ 
d\xi ^{*}q &  & q\geq (1-\xi ^{*})^{-1},
\end{array}
\right. 
\end{equation}
where $\xi ^{*}=\min_i\xi _i$. We see here that the function $\tau (q)$
``overlooks'' all the singularities $\xi _i$, except for the smallest one.
Therefore, any perturbation of the singularities $\xi _i$ that keeps $\xi^{*}$
invariant will change $f(\alpha )$ but not $\tau (q)$, which implies
that the mapping of $\tau (q)$ to $f(\alpha )$ must be indeterminate as
there is an infinite number of spectra associated with the same free energy. 
Physically, this also implies that $\tau(q)$ does not offer the most complete 
description of the dynamical system, since this function overlooks, as we said, all but 
one singularity. To really obtain a complete picture of all the singularities of the
system, one must resort to calculate $f(\alpha)$ and not just $\tau(q)$.

\section{Theory of LF transforms}
\label{sThLF}

The results of the two previous examples are very simple and show at once
that $f(\alpha )$ cannot in general be expressed as the LF\ transform of 
$\tau (q)$, contrary to what is claimed in most if not all references on the
subject. The problem, as was mentioned, is that LF transforms can only yield
concave functions, which means that these transforms cannot be used to
calculate nonconcave multifractal spectra, including those of the two
examples considered before. To make this observation more rigorous, we
introduce in this section a few concepts and results of convex analysis, 
beginning with the concept of supporting lines.
(All the definitions and theorems discussed here can be found in Ref.~\cite{rockafellar1970}; 
see also Chapter VI of Ref.~\cite{ellis1985} and Appendix A of Ref.~\cite{costeniuc2005}.)

\begin{definition}
\label{def1}
A function $f:\Bbb{R} \rightarrow \Bbb{R}$ admits a supporting line at $\alpha $ if there exists a
constant $\eta $ such that 
\begin{equation}
f(\beta )\leq f(\alpha )+\eta (\beta -\alpha )  \label{suppl1}
\end{equation}
for all $\beta \in \Bbb{R}$.
\end{definition}

\begin{figure}[t]
\centering
\includegraphics{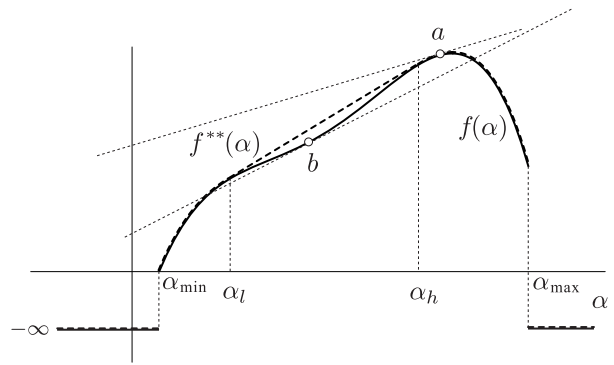}
\caption{(a) A generic nonconcave multifractal spectrum $f(\protect\alpha )$ 
(full line) together with its concave envelope $f^{**}(\protect\alpha )$ (dashed line).
The two functions coincide outside the
open interval $(\protect\alpha_l , \protect\alpha_h )$. The point $a$ 
of the multifractal spectrum admits a 
supporting line (concave point), while the point $b$ does not (nonconcave point).}
\label{ncs1}
\end{figure}

This definition means graphically that we can draw a line on top of the graph
of $f(\alpha )$ that does not go under that graph (see Fig.~\ref{ncs1}); hence the word ``supporting.''
With this picture in mind, it is easily seen that, if $f$ admits a
supporting line at $\alpha $ and is differentiable at $\alpha $, then the
slope $\eta $ of the supporting line must be such that $f^{\prime }(\alpha)=\eta $.

The importance of supporting lines comes from their association with LF
transforms, and from the fact, more precisely, that they determine whether
such transforms are \textit{involutive}, that is, whether they are their own
inverse. In the context of $f(\alpha )$ and $\tau (q)$, this means precisely
the following. First, recall that $\tau (q)$ can always be expressed as
the LF transform of $f(\alpha )$, so Eq.(\ref{LF1}) is always valid
independently of the shape of $f(\alpha )$. This follows essentially from
the fact that $\tau (q)$ is an always concave function of $q$ \cite{beck1993},
a fact that can be proved using H\"{o}lder's inequality. The inverse
transform shown in (\ref{LF2}), however, is not generally valid, and this is
where supporting lines become important, as expressed in the next theorem.

\begin{theorem}
\label{thmsl1}
If $f$ admits a supporting line at $\alpha $, then $f$ at $\alpha $ can expressed as the LF
transform of $\tau (q)$ as in Eq.(\ref{LF2}). In this case, we say that $f$
is \textit{concave} at $\alpha $. On the other hand, if $f$ does not admit a
supporting line at $\alpha $, then $f$ at $\alpha $ does not equal the LF
transform of $\tau (q)$. In this case, we say that $f$ is \textit{nonconcave}
at $\alpha $.
\end{theorem}

The two complementary results expressed in the theorem above are usually
rephrased in convex analysis by defining the function 
\begin{equation}
f^{**}(\alpha )=\inf_{q\in \Bbb{R}}\{q\alpha -\tau (q)\}.  \label{fconc1}
\end{equation}
In terms of $f^{**}(\alpha )$, we then have the following result.

\begin{theorem}
\label{thsip2}
$f(\alpha)=f^{**}(\alpha)$ if and only if $f$ admits a supporting
line at $\alpha$.
\end{theorem}

For the remaining, it is useful to note that $f^{**}(\alpha )$ corresponds
in general to the smallest concave function satisfying $f(\alpha )\leq
f^{**}(\alpha )$ for all $\alpha \in \Bbb{R}$. For this reason, 
$f^{**}(\alpha )$ is called the \textit{concave envelope} or \textit{concave
hull} of $f(\alpha )$. This implies, in particular, that if $f(\alpha )$
admits no supporting lines over some open interval, say $(\alpha _l,\alpha_h)$
as in Fig.~\ref{ncs1}, then $f^{**}(\alpha )$ must be \textit{affine}
over that interval, by which we mean that $f^{**}(\alpha )$ has a constant
slope over that interval. This last property, which is related to the
Maxwell construction \cite{huang1987,touchette2003}, is illustrated in
Fig.~\ref{ncs1}.

All of the properties of $f(\alpha )$ and $f^{**}(\alpha )$ in relation to
LF transforms can be verified for the two examples considered previously. In
the case of the invariant density of the Ulam map, for example, the concave
hull of $f(\alpha )$ is the function displayed in (\ref{finf1}); it is
obviously such that $f(\alpha )\leq f^{**}(\alpha )$ and is concave contrary
to $f(\alpha )$. Moreover, it is easily verified from Fig.~\ref{exfig1} that
the two points $\alpha =1/2$ and $\alpha =1$ admit a supporting line, which
explains why $f(\alpha )=f^{**}(\alpha )$ there. These two points admit in
fact an infinite number of supporting lines. For the point $\alpha =1/2$,
for example, all lines attached to $(1/2,0)$ with slope in the interval 
$[2,\infty )$ are supporting in the sense of (\ref{suppl1}). For $\alpha =1$,
the supporting lines have slopes in the interval $(-\infty ,2]$.

We can go further in our analysis of $f(\alpha )$ and $\tau (q)$ by calling
attention to the fact that
\begin{equation}
\tau (q)=\inf_{\alpha \in \Bbb{R}}\{q\alpha -f^{**}(\alpha )\}.
\end{equation}
Therefore, $\tau (q)$ is not only the LF transform of $f(\alpha )$, as
stated in Eq.(\ref{LF1}), but also the LF\ transform of $f^{**}(\alpha )$.
This result is general: it holds for any function $f(\alpha )$ and its
concave envelope $f^{**}(\alpha )$ defined as in Eq.(\ref{fconc1}) as the
double LF\ transform of $f(\alpha )$ or, more compactly, as 
\begin{equation}
f^{**}=\tau ^{*}=(f^{*})^{*},
\end{equation}
where the star stands for the LF transformation. To summarize, we then have 
$\tau =f^{*}$, $\tau ^{*}=(f^{*})^{*}=f^{**}$ and $(\tau
^{*})^{*}=(f^{**})^{*}=f^{*}=\tau $. This chain of equalities can be
expressed in a more transparent way using the following diagram: 
\begin{equation}
f(\alpha )\stackrel{*}{\rightarrow }\tau (q)\stackrel{*}{\rightleftharpoons }
f^{**}(\alpha ),  \label{diag1}
\end{equation}
which makes obvious the fact that there may be more than one spectrum
related the same free energy. In fact, all $f(\alpha )$ having the same
concave envelope lead to the same $\tau (q)$, as can be verified in the
second example considered before. Finally, note that the chain of equalities
reduces to $\tau =f^{*}$ and $\tau ^{*}=f$, or equivalently to 
\begin{equation}
\tau (q)\stackrel{*}{\rightleftharpoons }f(\alpha ),  \label{diag2}
\end{equation}
when $f(\alpha )=f^{**}(\alpha )$ for all $\alpha \in \Bbb{R}$, that is,
when $f(\alpha )$ is everywhere concave.

\begin{figure}[t]
\centering
\includegraphics{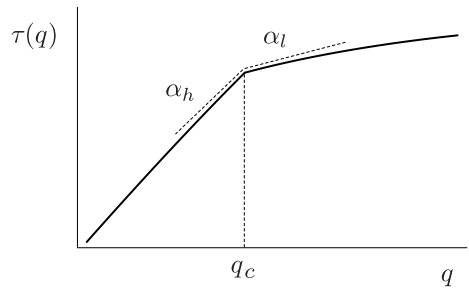}
\caption{Free energy function $\protect\tau (q)$
associated with the multifractal spectrum
$f(\protect\alpha )$ shown in Fig.~\protect\ref{ncs1}.
The LF transform of the concave envelope $f^{**}(\protect\alpha)$ of 
$f(\protect\alpha )$ yields the same free energy function.}
\label{tau1}
\end{figure}

Having listed all the relationships that exist between $f(\alpha )$, $\tau(q)$
and $f^{**}(\alpha )$, we can now fully address the main issue of this
paper, which is to determine when $f(\alpha )$ can safely and completely be
calculated as the LF transform of $\tau (q)$. From the chain of equalities
and diagrams shown above, this amounts to determine when the LF transform is
involutive; that is to say, under which conditions does the diagram (\ref{diag1})
reduce to the diagram of (\ref{diag2})?

A first obvious answer to this question is given by recalling what we have
just mentioned about the diagram of (\ref{diag2}), namely that \textit{if}
$f(\alpha )$\textit{\ is everywhere concave,} \textit{then the multifractal
spectra $f(\alpha )$ can completely be calculated as the LF
transform of the free energy function $\tau (q)$}.  As such, this
answer is complete but not very practical because it is based on $f(\alpha )$
and so presupposes that we know $f(\alpha )$. A more useful criterion can be
stated from the point of view of $\tau (q)$ alone by using a
result of convex analysis connecting nonconcave or affine regions of
$f(\alpha )$ with nondifferentiable points of $\tau (q)$. This result is stated next without
a proof; see \cite{rockafellar1970} for more details (see especially Theorems 23.5 and 26.3).

\begin{theorem}
\label{thmt1}
Suppose that $f(\alpha )$ is nonconcave over some open interval $(\alpha _l,\alpha _h)$
(Fig.~\ref{ncs1}) or that $f(\alpha )$ is concave but affine over $(\alpha _l,\alpha
_h)$. Then $\tau (q)$ is nondifferentiable at some critical value $q_c$
corresponding to the slope of $f^{**}(\alpha )$ over the interval $(\alpha
_l,\alpha _h)$. Moreover, the left- and right-derivatives of $\tau (q)$ at
$q_c$ equal $\alpha _h$ and $\alpha _l$, respectively (Fig.~\ref{tau1}).
\end{theorem}

From this result, we arrive at our criterion by taking the contrapositive:
if $\tau (q)$ is everywhere differentiable, then $f(\alpha )$ is concave
everywhere with no affine parts. Thus, from the point of view of $\tau (q)$, 
$f(\alpha )$\textit{can completely be calculated as the LF transform of
$\tau (q)$ if the latter function is everywhere differentiable}. 
Taking the view that nondifferentiable points of $\tau (q)$ represent
first-order phase transitions for multifractals
\cite{tremblay1986,fourcade1987,szepfalusy1987,csordas1989,jensen1988},
this is equivalent to saying that $f(\alpha )$ \textit{can completely be calculated as the
LF transform of $\tau (q)$ in the absence of first-order phase
transitions}. If there is a first-order phase transition, then either
$f(\alpha )$ is nonconcave somewhere, in which case $f\neq \tau ^{*}$, or
else $f(\alpha )$ is affine somewhere, in which case $f=\tau ^{*}$.
Unfortunately---and this is an important point---there is no way to
distinguish the two cases from the sole knowledge of $\tau (q)$
(see Figs.~\ref{ncs1} and \ref{tau1}). Thus, if $\tau (q)$ has one or more
nondifferentiable points and if there is no reason to think that $f(\alpha )$
is concave, then one must resort to calculate $f(\alpha )$ by means which do
not rely on $\tau (q)$.

\section{Applications}
\label{sApp}

We now revisit some examples of multifractal models that have been
discussed in the physics literature, and point out where and how our results
of the previous section apply.

\subsection{Multifractal turbulence models}

One of the first field of study for which multifractal models have been
developed is fully developed hydrodynamic turbulence
\cite{frisch1985,benzi1991}. The basis of these models is that, in the turbulent flow,
velocity increments $\delta v(l)=|v(x+l)-v(x)|$ at a given distance $l$
scale with local exponents $h$, which are distributed on a fractal set with
fractal dimension $D(h)$. This notation is taken from Benzi 
\textit{et al}.~\cite{benzi1991}, and can be translated to our notation using the following
identifications: 
\begin{eqnarray}
h &=&\alpha   \nonumber \\
l &=&\varepsilon   \nonumber \\
3-D(h) &=&-f(\alpha )  \nonumber \\
p &=&q  \nonumber \\
\zeta _p &=&\tau (q).
\end{eqnarray}
Here $\zeta _p$ denote the scaling exponents of moments of velocity
increments in the inertial range, 
\begin{equation}
\langle (\delta v)^p\rangle \sim l^{\zeta _p}.
\end{equation}
In multifractal turbulence models, the probability to observe a local
exponent $h$ is given by 
\begin{equation}
P_l(h)\sim l^{3-D(h)},
\end{equation}
which is equivalent to our notation $n_\varepsilon (\alpha )\sim \varepsilon^{-f(\alpha )}$.
Moreover, for the scaling exponents $p$ one has 
\begin{equation}
\zeta _p=\min_h\{hp+3-D(h)\},
\end{equation}
which is equivalent to the LF\ transform of Eq.(\ref{LF1}). In practice, one
extracts $D(h)$ from the scaling exponents $\zeta _p$, which can be measured
in experiments. In our notation, this means that one determines $f(\alpha )$
by the LF transform of the experimentally measured $\tau (q)$. 

In view of all the results derived before, we can notice that multifractal
turbulence models in their current form can only deal with the convex hull
of the spectrum of singularities. The true spectrum $D(h)$ of a turbulent
flow is fully determined by the underlying dynamics, i.e., the Navier-Stokes
equation, and there is a priori no reason to think that this spectrum should be a
concave function of $h$. Our arguments of the previous section now allow for
an experimental check of the concavity of $D(h)$: if the
experimentally-measured $\zeta _p$ is differentiable within the precision
allowed by the experiment, then $D(h)$ is concave and therefore given by
the LF transform of $\zeta _p$. If one observes that $\zeta _p$ is
nondifferentiable, then either $D(h)$ is nonconcave or else is affine. Both cases are consistent with the fact that $\zeta_p$ is nondifferentiable, but there is no way to tell from $\zeta_p$ which one of the two spectra is the actual one.

The exponents $\zeta _p$ have been measured in many experiments; see, e.g.,
Refs.~\cite{benzi1993,stolovitzky1993}. Within the experimental
uncertainties, they are usually described by a smooth function of $p$,
although one cannot fully exclude the existence of phase transitions. The
data found in Refs.~\cite{benzi1993,stolovitzky1993}, for example, show a
relatively strong change of slope near $p=3$. If the existence of such
transitions were confirmed (e.g., via the study of theoretical models of 
turbulence), then one would have to check that the underlying
spectrum $D(h)$ is either nonconcave or concave but affine somewhere. If $D(h)$
is affine (see, e.g., Ref.~\cite{xu2006}), then that spectrum is correctly given by the LF transform of $\zeta_p$. If $D(h)$ is nonconcave, then that spectrum
is not fully given by the LF transform of $\zeta_p$. In this case, $\zeta_p$
cannot provide a complete description of the turbulent flow, since all the 
local exponents $h$ in the nonconcave region of $D(h)$ are ``overlooked,'' 
in the spirit of the two simple examples discussed before, by $\zeta_p$. 
In short, these exponents are described by $D(h)$ but not by $\zeta_p$.

\subsection{Diffusion-limited aggregates}

For the second example, we consider multifractals as generated by
diffusion-limited aggregates (DLA) \cite{witten1981,halsey21986,jensen2002}. Jensen 
\textit{et al.}~\cite{jensen2002} provide convincing evidence that the
function $\tau (q)$ calculated for the harmonic measure of their DLA cluster
exhibits a first-order phase transition at $q=-0.23\pm 0.05$ (see their
Fig.~3). From this result, all that can be said about their $f(\alpha )$
spectrum obtained by LF-transforming $\tau (q)$ (shown in their Fig.~4) is
that it is the concave hull $f^{**}(\alpha )$ of the true $f(\alpha )$
spectrum. Note that the spectrum displayed in Fig.~4 of Ref.~\cite{jensen2002}
does show an affine part, so it is consistent with the fact
that $\tau (q)$ has a nondifferentiable point. However, there is a priori no
reason why DLA clusters should possess a concave spectrum of singularities,
so that the part where $f(\alpha )$ is seen to be affine could just as well
be nonconcave. Therefore, at this point we may conclude that the true 
$f(\alpha )$ spectrum of the DLA cluster studied by Jensen \textit{et al}.~is
as yet unknown.

It should be remarked that, although Jensen \textit{et al}.~provide
evidence to the effect that $\tau (q)$ possesses a nondifferentiable
point, the $\tau (q)$ which is calculated in practice is actually always
analytic if one deals with finite-size DLA\ clusters. The spectrum
$f^{**}(\alpha )$ which is calculated from $\tau (q)$ is, in this case,
necessarily concave and has no affine parts. The nondifferentiable point of 
$\tau (q)$ and the concomitant affine part of $f^{**}(\alpha )$ appear,
formally speaking, only in the ``thermodynamic'' limit of infinitely large
clusters.

\subsection{Chaotic systems}

Multifractal spectra have been calculated for many examples of chaotic dynamical systems, including the H\'enon map \cite{hata1989} (see also Ref.~\cite{mori1989}), and the driven damped pendulum \cite{tominaga1990}. The particularity of these two examples is that they seem to give rise to first-order phase transitions, which are also referred to as $q$-phase transitions. Accordingly, the question arises as to whether these phase transitions emerge out of a nonconcave $f(\alpha)$ or an affine $f(\alpha)$.

The question, as it stands, is not resolved in the papers that treat these examples because they assume that $f(\alpha)$ is always the LF transform of $\tau(q)$, which means that they implicitly assume that $f(\alpha)$ is always concave. It must be observed that some of the reported spectra appear to be affine (see, e.g., Fig.~2 in Ref.~\cite{hata1989}), so they are not problematic---they satisfy the concavity assumption. However, some other spectra are clearly nonconcave; see, e.g., Fig.~1 in Ref.~\cite{hata1989} and Figs.~2 and 3 in Ref.~\cite{tominaga1990}. For these, we must be careful because the observed nonconcavity could be a finite-size effect inherent to the fact that $f(\alpha)$ is computed numerically for a finite coarse-graining resolution $\varepsilon$. Thus it could be that the nonconcavity of $f(\alpha)$ observed for $\varepsilon > 0 $ disappears as $\varepsilon\rightarrow 0$. To verify this, one would need to perform a finite-size analysis of the data by computing $f(\alpha)$ for 
decreasing values of $\varepsilon$ and study the convergence of the results.
Similar finite-size analyses have been performed in the context of numerical
calculations of the microcanonical entropy function; 
see Refs.~\cite{kastner2000,pleimling2004,behringer2005}.

\subsection{Spectrum of dynamical indices}

First-order phase transitions have also been studied in the context of chaotic systems at the level of an entropy-like quantity referred to as the spectrum of dynamical indices or expansion-rate spectrum \cite{sano1986,hata1988,mori1989,beck1993}. These phase transitions typically occur for nonhyperbolic dynamical systems. 

It would take us too far to explain the notion of dynamical indices and the many examples for which this quantity has been studied. Let us only mention that the spectrum of dynamical indices is a dynamical analog of $f(\alpha)$, and that affine and nonconcave spectra of dynamical indices have been reported in the literature; see, e.g., Refs.~\cite{horita1988,hata1988, tomita1988,tomita1989,yoshida1989,mori1989}. 
Most of these references, unfortunately, share the same problem as those discussed so far: they assume that the spectrum of dynamical indices can always be calculated as the LF transform of a dynamical analogue of the free energy function $\tau (q)$. Most of them assume this even when reporting the computation of nonconcave spectra; see Yoshida and Miyazaki \cite{yoshida1989} for a noticeable exception. 

\section{Generalized free energy functions}

At this point, we have emphasized more than once that a nonconcave multifractal spectrum cannot be calculated as the LF transform of its corresponding free energy function. Our goal in this section is to offer a practical solution to this problem by illustrating a method for obtaining nonconcave spectra through LF transforms of a \textit{generalized} form of free energy function. The method was proposed recently in  Refs.~\cite{costeniuc2005,costeniuc2006} in the context of nonconcave entropies within the microcanonical ensemble, and will be illustrated here in the context of the
first example considered in Sec.~\ref{sExamples}.

The form of generalized free energy that we shall consider is based on a generalization of the partition function given by 
\begin{equation}
Z_\varepsilon (q,g)=\sum_i\varepsilon ^{q\alpha _i+g(\alpha _i)},
\end{equation}
where $\alpha _i$ represents the local fractal exponent associated with the
probability $p_{\varepsilon ,i}$, and $g$ is an arbitrary smooth function.
This new form generalizes the standard canonical partition function, in the
sense that 
\begin{equation}
Z_\varepsilon (q,g=0)=Z_\varepsilon (q)=\sum_i\varepsilon ^{q\alpha _i}.
\end{equation}
For definiteness, we shall adopt the choice $g(\alpha )=\gamma \alpha ^2$
with $\gamma \in \Bbb{R}$. Therefore, the generalized partition function
that we consider is 
\begin{equation}
Z_\varepsilon (q,\gamma )=\sum_i\varepsilon ^{q\alpha _i+\gamma \alpha _i^2}.
\end{equation}
We call this partition function the \textit{Gaussian partition function}; its
corresponding free energy function 
\begin{equation}
\tau (q,\gamma )=\lim_{\varepsilon \rightarrow 0}\frac{\ln Z_\varepsilon
(q,\gamma )}{\ln \varepsilon }
\end{equation}
is called the \textit{Gaussian free energy}. Note that this new free energy
is a function of two real parameters, $q$ and $\gamma $, and that $\tau
(q,\gamma =0)=\tau (q)$.

The rationale for generalizing the standard free energy function $\tau (q)$
to $\tau (q,\gamma )$ is that it modifies the structure of the LF transform
which connects $\tau (q)$ with $f(\alpha )$, and thus modifies the
conditions which ensure that $f(\alpha )$ can be written as the LF transform
of a free energy function. We spare the reader with the details of this
modification which can be found in Refs.~\cite{costeniuc2005,costeniuc2006}.
For our purpose, we shall only note the generalized versions of the LF
transforms that connect $\tau (q,\gamma )$ and $f(\alpha )$; they are given by 
\begin{equation}
\tau (q,\gamma )=\inf_\alpha \{q\alpha +\gamma \alpha ^2-f(\alpha )\}
\label{gLF1}
\end{equation}
and 
\begin{equation}
f(\alpha )=\inf_{q,\gamma }\{q\alpha +\gamma \alpha ^2-\tau (q,\gamma )\}.
\label{gLF2}
\end{equation}
The first LF transform holds, like its standard version ($\gamma =0$), for
any spectrum $f(\alpha )$, be it concave or not. The surprising virtue of
the second LF\ transform is that it also holds true for basically any 
$f(\alpha )$, contrary to the standard version ($\gamma =0$) which applies
only when $f(\alpha )$ is concave. Rather than proving this result, we shall
verify that it is valid for the nonconcave multifractal spectrum shown in
Fig.~\ref{exfig1}(a). That is, we shall obtain that nonconcave spectrum by
inverting, in the manner of Eq.(\ref{gLF2}), its associated Gaussian free
energy $\tau (q,\gamma )$.

First, we calculate $\tau (q,\gamma )$ starting from $Z_\varepsilon
(q,\gamma )$: 
\begin{equation}
Z_\varepsilon (q,\gamma )=\sum_i\varepsilon ^{q\alpha _i+\gamma \alpha
_i^2}\sim \varepsilon ^{q/2+\gamma /4}+\varepsilon ^{-1}\varepsilon
^{q+\gamma }.
\end{equation}
Taking the limit $\varepsilon \rightarrow 0$ yields 
\begin{equation}
\tau (q,\gamma )=\min \{q/2+\gamma /4,q+\gamma -1\}.
\end{equation}
The solution of the minimum can be found explicitly; it has the form 
\begin{equation}
\tau (q,\gamma )=\left\{ 
\begin{array}{lll}
q/2+\gamma /4 &  & q\geq q_\gamma  \\ 
q+\gamma -1 &  & q<q_\gamma ,
\end{array}
\right. 
\end{equation}
where $q_\gamma =-3\gamma /2+2$. Next, we apply formula (\ref{gLF2}) using
this solution for $\tau (q,\gamma )$. This leads us to solving the following
variational problem: 
\begin{equation}
I=\inf_{q,\gamma }\left\{ 
\begin{array}{lll}
q\alpha +\gamma \alpha ^2-q/2-\gamma /4 &  & q\geq q_\gamma  \\ 
q\alpha +\gamma \alpha ^2-q-\gamma +1 &  & q<q_\gamma 
\end{array}
\right\} .
\end{equation}
Grouping the variables together, this is equivalent to 
\begin{equation}
I=\inf_{q,\gamma }\left\{ 
\begin{array}{lll}
q(\alpha -1/2)+\gamma (\alpha ^2-1/4) &  & q\geq q_\gamma  \\ 
q(\alpha -1)+\gamma (\alpha ^2-1)+1 &  & q<q_\gamma 
\end{array}
\right\} .
\end{equation}
Let $f_1(\alpha ,q,\gamma )$ denote the top expression in the brackets and
$f_2(\alpha ,q,\gamma )$ the lower one. With this notation, it can be noted
that, for $\alpha =1/2$, 
\begin{equation}
f_2(1/2,q,\gamma )>f_1(1/2,q,\gamma )=0
\end{equation}
for all $q<q_\gamma $ and $\gamma \in \Bbb{R}$. Therefore, 
\begin{equation}
\inf_{q<q_\gamma ,\gamma }f_2(\alpha ,q,\gamma )=f_1=0,
\end{equation}
and $I=0$ at $\alpha =1/2$. Similarly, for $\alpha =1$, we have 
\begin{equation}
f_1(1,q,\gamma )\geq f_2(1,q,\gamma )=1
\end{equation}
for all $q\geq q_\gamma $ and $\gamma \in \Bbb{R}$, so that $I=f_2=1$ at
$\alpha =1$. For all other values of $\alpha $, it is possible to set $q$ and 
$\gamma $ in such a way that $I=-\infty $. At the end, we are left with
\begin{equation}
I=\left\{ 
\begin{array}{lll}
0 &  & \alpha =1/2 \\ 
1 &  & \alpha =1 \\ 
-\infty  &  & \text{otherwise,}
\end{array}
\right. 
\end{equation}
which is the precise expression of $f(\alpha)$, as
given in Eq.(\ref{initf1}). Therefore, we have shown that this
nonconcave spectrum can be expressed as in Eq.(\ref{gLF2}) as a modified LF
transform of a generalized free energy. 

The same method can be applied to calculate other nonconcave spectra. In fact, it has been conjectured that the method can be used to calculate \textit{any} nonconcave spectrum (viz., nonconcave entropy function) as the LF transform of a properly-chosen generalized free energy. More details about this universality property can be found in Refs.~\cite{costeniuc2005,costeniuc2006}.

\section{Conclusion}

We have shown in this paper that one must be careful when calculating the
singularity spectrum of multifractals as the LF transform of its
corresponding free energy, since LF transforms can only yield concave
functions. This word of caution has implications for most of the studies published
so far on multifractals, including those on multifractal models of
turbulence, as they have taken for granted that the multifractal
spectrum is the LF transform of the free energy no matter what the spectrum
looks like. This, as we have seen, is only true if the spectrum is concave;
if it is nonconcave, then one must resort to calculate it directly from its
definition. Another possibility is to use a generalization of
the canonical ensemble which can be used to extract nonconcave entropies
from a generalized version of the free energy function. This way of doing
was sketched here in the context of a simple example of nonconcave
multifractal spectrum, and is presented in full details in
Refs.~\cite{costeniuc2005,costeniuc2006}.

In the end, it should be noted that the results that we have discussed in this paper are not special to multifractals, but apply actually to any field of investigation which uses LF transforms. In statistical mechanics, for example, the LF transform that connects the
entropy function of the microcanonical ensemble with the free energy
function of the canonical ensemble becomes noninvolutive when the entropy is
nonconcave. When this happens, we say that there is nonequivalence of
ensembles \cite{ellis2000,touchette2004}, since one is then unable to obtain
the true entropy function of the microcanonical ensemble solely from the
knowledge of the free energy of the canonical ensemble. The notion of
generalized canonical ensemble has been developed precisely in this context.

Similarly, in large deviation theory, it has been known for some time that
nonconvex rate functions cannot be calculated by means of LF transforms of functions analogous to free energy functions \cite{dinwoodie1992,ioffe1993,ellis1995,touchette22003}. A multifractal spectrum is in essence an entropy function, and an entropy function is in essence a rate function \cite{plachky1975,lanford1973,zohar1999,veneziano2002}, so there is actually a deep connection with what we have presented here and what is known in large deviation theory \cite{touchette22003}. For example, the result relating the differentiability of $\tau(q)$ and the (strict) concavity of $f(\alpha)$ can be put in correspondence with a result of large deviation theory known as the G\"artner-Ellis Theorem \cite{ellis1985,oono1989,ellis1995}. Furthermore, the result stating that the LF transform of $f(\alpha)$ always yields $\tau(q)$ (see Sec.~\ref{sThLF}) can be put in correspondence with a large deviation result known as Varadhan Theorem; see Refs.~\cite{ellis1985,oono1989}.

From these correspondences, it is but a small step to conjecture that
nonconcave entropies or, more generally, nonconvex rate functions should show up in other physical theories in which large deviations are at play. One such theory that comes to mind is the thermodynamic formalism of dynamical systems \cite{ruelle2004,beck1993}; another is nonequilibrium statistical mechanics. We have already alluded to the first theory when discussing the spectrum of dynamical indices in Sec.~\ref{sApp}. Concerning nonequilibrium statistical mechanics, the reader will find an example of nonconvex rate function in a recent paper by Imparato and Peliti \cite{imparato2005} (see their Fig.~12). This paper discusses thermodynamic fluctuations in systems driven out of equilibrium. A large deviation result in that context is referred to as a fluctuation theorem, whereas a rate function is called a fluctuation function.

\section*{Addendum}

Since this paper was submitted for publication, we became aware of two recent papers by Testud \cite{testud2005,testud2006} and one by Riedi \cite{riedi1995} in which examples of nonconcave multifractal spectra are discussed.

\begin{acknowledgments}
H.T. was supported by NSERC (Canada), the Royal Society of London, and FCAR
(Qu\'{e}bec) during the initial phase of this work. C.B. is supported by a Springboard
Fellowship from EPSRC (UK).
\end{acknowledgments}

\bibliography{multifract}

\begin{thebibliography}{60}
\expandafter\ifx\csname natexlab\endcsname\relax\def\natexlab#1{#1}\fi
\expandafter\ifx\csname bibnamefont\endcsname\relax
  \def\bibnamefont#1{#1}\fi
\expandafter\ifx\csname bibfnamefont\endcsname\relax
  \def\bibfnamefont#1{#1}\fi
\expandafter\ifx\csname citenamefont\endcsname\relax
  \def\citenamefont#1{#1}\fi
\expandafter\ifx\csname url\endcsname\relax
  \def\url#1{\texttt{#1}}\fi
\expandafter\ifx\csname urlprefix\endcsname\relax\def\urlprefix{URL }\fi
\providecommand{\bibinfo}[2]{#2}
\providecommand{\eprint}[2][]{\url{#2}}

\bibitem[{\citenamefont{Falconer}(1990)}]{falconer1990}
\bibinfo{author}{\bibfnamefont{K.}~\bibnamefont{Falconer}},
  \emph{\bibinfo{title}{Fractal Geometry: Mathematical Foundations and
  Applications}} (\bibinfo{publisher}{Wiley}, \bibinfo{address}{New York},
  \bibinfo{year}{1990}).

\bibitem[{\citenamefont{Beck and Schl{\"o}gl}(1993)}]{beck1993}
\bibinfo{author}{\bibfnamefont{C.}~\bibnamefont{Beck}} \bibnamefont{and}
  \bibinfo{author}{\bibfnamefont{F.}~\bibnamefont{Schl{\"o}gl}},
  \emph{\bibinfo{title}{Thermodynamics of Chaotic Systems}}
  (\bibinfo{publisher}{Cambridge University Press},
  \bibinfo{address}{Cambridge}, \bibinfo{year}{1993}).

\bibitem[{\citenamefont{Mandelbrot}(1999)}]{mandelbrot1999}
\bibinfo{author}{\bibfnamefont{B.~B.} \bibnamefont{Mandelbrot}},
  \emph{\bibinfo{title}{Multifractals and $1/f$ Noise}}
  (\bibinfo{publisher}{Springer}, \bibinfo{address}{New York},
  \bibinfo{year}{1999}).

\bibitem[{\citenamefont{Frisch and Parisi}(1985)}]{frisch1985}
\bibinfo{author}{\bibfnamefont{U.}~\bibnamefont{Frisch}} \bibnamefont{and}
  \bibinfo{author}{\bibfnamefont{G.}~\bibnamefont{Parisi}}, in
  \emph{\bibinfo{booktitle}{Turbulence and Predictibility of Geophysical Flows
  and Climate Dynamics}}, edited by
  \bibinfo{editor}{\bibfnamefont{M.}~\bibnamefont{Ghil}},
  \bibinfo{editor}{\bibfnamefont{R.}~\bibnamefont{Benzi}}, \bibnamefont{and}
  \bibinfo{editor}{\bibfnamefont{G.}~\bibnamefont{Parisi}}
  (\bibinfo{publisher}{North-Holland}, \bibinfo{address}{Amsterdam},
  \bibinfo{year}{1985}).

\bibitem[{\citenamefont{Halsey et~al.}(1986{\natexlab{a}})\citenamefont{Halsey,
  Jensen, Kadanoff, Procaccia, and Shraiman}}]{halsey1986}
\bibinfo{author}{\bibfnamefont{T.~C.} \bibnamefont{Halsey}},
  \bibinfo{author}{\bibfnamefont{M.~H.} \bibnamefont{Jensen}},
  \bibinfo{author}{\bibfnamefont{L.~P.} \bibnamefont{Kadanoff}},
  \bibinfo{author}{\bibfnamefont{I.}~\bibnamefont{Procaccia}},
  \bibnamefont{and} \bibinfo{author}{\bibfnamefont{B.~I.}
  \bibnamefont{Shraiman}}, \bibinfo{journal}{Phys. Rev. A}
  \textbf{\bibinfo{volume}{33}}, \bibinfo{pages}{1141}
  (\bibinfo{year}{1986}{\natexlab{a}}).

\bibitem[{\citenamefont{Benzi et~al.}(1984)\citenamefont{Benzi, Paladin,
  Parisi, and Vulpiani}}]{benzi1984}
\bibinfo{author}{\bibfnamefont{R.}~\bibnamefont{Benzi}},
  \bibinfo{author}{\bibfnamefont{G.}~\bibnamefont{Paladin}},
  \bibinfo{author}{\bibfnamefont{G.}~\bibnamefont{Parisi}}, \bibnamefont{and}
  \bibinfo{author}{\bibfnamefont{A.}~\bibnamefont{Vulpiani}},
  \bibinfo{journal}{J. Phys. A} \textbf{\bibinfo{volume}{17}},
  \bibinfo{pages}{3521} (\bibinfo{year}{1984}).

\bibitem[{\citenamefont{Benzi et~al.}(1991)\citenamefont{Benzi, Biferale,
  Paladin, Vulpiani, and Vergassola}}]{benzi1991}
\bibinfo{author}{\bibfnamefont{R.}~\bibnamefont{Benzi}},
  \bibinfo{author}{\bibfnamefont{L.}~\bibnamefont{Biferale}},
  \bibinfo{author}{\bibfnamefont{G.}~\bibnamefont{Paladin}},
  \bibinfo{author}{\bibfnamefont{A.}~\bibnamefont{Vulpiani}}, \bibnamefont{and}
  \bibinfo{author}{\bibfnamefont{M.}~\bibnamefont{Vergassola}},
  \bibinfo{journal}{Phys. Rev. Lett.} \textbf{\bibinfo{volume}{67}},
  \bibinfo{pages}{2299} (\bibinfo{year}{1991}).

\bibitem[{\citenamefont{Benzi et~al.}(1993)\citenamefont{Benzi, Ciliberto,
  Tripiccione, Baudet, Massaioli, and Succi}}]{benzi1993}
\bibinfo{author}{\bibfnamefont{R.}~\bibnamefont{Benzi}},
  \bibinfo{author}{\bibfnamefont{S.}~\bibnamefont{Ciliberto}},
  \bibinfo{author}{\bibfnamefont{R.}~\bibnamefont{Tripiccione}},
  \bibinfo{author}{\bibfnamefont{C.}~\bibnamefont{Baudet}},
  \bibinfo{author}{\bibfnamefont{F.}~\bibnamefont{Massaioli}},
  \bibnamefont{and} \bibinfo{author}{\bibfnamefont{S.}~\bibnamefont{Succi}},
  \bibinfo{journal}{Phys. Rev. E} \textbf{\bibinfo{volume}{48}},
  \bibinfo{pages}{29} (\bibinfo{year}{1993}).

\bibitem[{\citenamefont{Stolovitzky et~al.}(1993)\citenamefont{Stolovitzky,
  Sreenivasan, and Juneja}}]{stolovitzky1993}
\bibinfo{author}{\bibfnamefont{G.}~\bibnamefont{Stolovitzky}},
  \bibinfo{author}{\bibfnamefont{K.~R.} \bibnamefont{Sreenivasan}},
  \bibnamefont{and} \bibinfo{author}{\bibfnamefont{A.}~\bibnamefont{Juneja}},
  \bibinfo{journal}{Phys. Rev. E} \textbf{\bibinfo{volume}{48}},
  \bibinfo{pages}{3212} (\bibinfo{year}{1993}).

\bibitem[{\citenamefont{Schertzer and Lovejoy}(1985)}]{schertzer1985}
\bibinfo{author}{\bibfnamefont{D.}~\bibnamefont{Schertzer}} \bibnamefont{and}
  \bibinfo{author}{\bibfnamefont{S.}~\bibnamefont{Lovejoy}},
  \bibinfo{journal}{Phys. Chem. Hydrodyn. J.} \textbf{\bibinfo{volume}{6}},
  \bibinfo{pages}{623} (\bibinfo{year}{1985}).

\bibitem[{\citenamefont{Schertzer et~al.}(1997)\citenamefont{Schertzer,
  Lovejoy, Schmitt, Chigirinskaya, and Marsan}}]{schertzer1997}
\bibinfo{author}{\bibfnamefont{D.}~\bibnamefont{Schertzer}},
  \bibinfo{author}{\bibfnamefont{S.}~\bibnamefont{Lovejoy}},
  \bibinfo{author}{\bibfnamefont{F.}~\bibnamefont{Schmitt}},
  \bibinfo{author}{\bibfnamefont{Y.}~\bibnamefont{Chigirinskaya}},
  \bibnamefont{and} \bibinfo{author}{\bibfnamefont{D.}~\bibnamefont{Marsan}},
  \bibinfo{journal}{Fractals} \textbf{\bibinfo{volume}{5}},
  \bibinfo{pages}{427} (\bibinfo{year}{1997}).

\bibitem[{\citenamefont{Schertzer and Lovejoy}(1987)}]{schertzer1987}
\bibinfo{author}{\bibfnamefont{D.}~\bibnamefont{Schertzer}} \bibnamefont{and}
  \bibinfo{author}{\bibfnamefont{S.}~\bibnamefont{Lovejoy}},
  \bibinfo{journal}{J. Geophys. Res.} \textbf{\bibinfo{volume}{92}},
  \bibinfo{pages}{9693} (\bibinfo{year}{1987}).

\bibitem[{\citenamefont{Schertzer and Lovejoy}(1989)}]{schertzer1989}
\bibinfo{author}{\bibfnamefont{D.}~\bibnamefont{Schertzer}} \bibnamefont{and}
  \bibinfo{author}{\bibfnamefont{S.}~\bibnamefont{Lovejoy}}, in
  \emph{\bibinfo{booktitle}{Fractals: Physical Origin and Consequences}},
  edited by \bibinfo{editor}{\bibfnamefont{L.}~\bibnamefont{Pietronero}}
  (\bibinfo{publisher}{Plenum}, \bibinfo{address}{New York},
  \bibinfo{year}{1989}), p.~\bibinfo{pages}{49}.

\bibitem[{\citenamefont{Schertzer and Lovejoy}(1991)}]{schertzer1991}
\bibinfo{editor}{\bibfnamefont{D.}~\bibnamefont{Schertzer}} \bibnamefont{and}
  \bibinfo{editor}{\bibfnamefont{S.}~\bibnamefont{Lovejoy}}, eds.,
  \emph{\bibinfo{title}{Scaling, Fractals and Non-Linear Variability in
  Geophysics}} (\bibinfo{publisher}{Kluwer}, \bibinfo{address}{Boston},
  \bibinfo{year}{1991}).

\bibitem[{\citenamefont{Schmitt et~al.}(1999)\citenamefont{Schmitt, Schertzer,
  and Lovejoy}}]{schmitt1999}
\bibinfo{author}{\bibfnamefont{F.}~\bibnamefont{Schmitt}},
  \bibinfo{author}{\bibfnamefont{D.}~\bibnamefont{Schertzer}},
  \bibnamefont{and} \bibinfo{author}{\bibfnamefont{S.}~\bibnamefont{Lovejoy}},
  \bibinfo{journal}{Appl. Stochastic Models Data Anal.}
  \textbf{\bibinfo{volume}{15}}, \bibinfo{pages}{29} (\bibinfo{year}{1999}).

\bibitem[{\citenamefont{Paladin and Vulpiani}(1987)}]{paladin1987}
\bibinfo{author}{\bibfnamefont{G.}~\bibnamefont{Paladin}} \bibnamefont{and}
  \bibinfo{author}{\bibfnamefont{A.}~\bibnamefont{Vulpiani}},
  \bibinfo{journal}{Phys. Rep.} \textbf{\bibinfo{volume}{156}},
  \bibinfo{pages}{147} (\bibinfo{year}{1987}).

\bibitem[{\citenamefont{Ott et~al.}(1984)\citenamefont{Ott, Withers, and
  Yorke}}]{ott1984}
\bibinfo{author}{\bibfnamefont{E.}~\bibnamefont{Ott}},
  \bibinfo{author}{\bibfnamefont{W.~D.} \bibnamefont{Withers}},
  \bibnamefont{and} \bibinfo{author}{\bibfnamefont{J.~A.} \bibnamefont{Yorke}},
  \bibinfo{journal}{J. Phys. Stat.} \textbf{\bibinfo{volume}{36}},
  \bibinfo{pages}{687} (\bibinfo{year}{1984}).

\bibitem[{\citenamefont{Rockafellar}(1970)}]{rockafellar1970}
\bibinfo{author}{\bibfnamefont{R.~T.} \bibnamefont{Rockafellar}},
  \emph{\bibinfo{title}{Convex Analysis}} (\bibinfo{publisher}{Princeton
  University Press}, \bibinfo{address}{Princeton}, \bibinfo{year}{1970}).

\bibitem[{\citenamefont{Ellis}(1985)}]{ellis1985}
\bibinfo{author}{\bibfnamefont{R.~S.} \bibnamefont{Ellis}},
  \emph{\bibinfo{title}{Entropy, Large Deviations, and Statistical Mechanics}}
  (\bibinfo{publisher}{Springer-Verlag}, \bibinfo{address}{New York},
  \bibinfo{year}{1985}).

\bibitem[{\citenamefont{Costeniuc et~al.}(2005)\citenamefont{Costeniuc, Ellis,
  Touchette, and Turkington}}]{costeniuc2005}
\bibinfo{author}{\bibfnamefont{M.}~\bibnamefont{Costeniuc}},
  \bibinfo{author}{\bibfnamefont{R.~S.} \bibnamefont{Ellis}},
  \bibinfo{author}{\bibfnamefont{H.}~\bibnamefont{Touchette}},
  \bibnamefont{and}
  \bibinfo{author}{\bibfnamefont{B.}~\bibnamefont{Turkington}},
  \bibinfo{journal}{J. Stat. Phys.} \textbf{\bibinfo{volume}{119}},
  \bibinfo{pages}{1283} (\bibinfo{year}{2005}), \eprint{cond-mat/0408681}.

\bibitem[{\citenamefont{Huang}(1987)}]{huang1987}
\bibinfo{author}{\bibfnamefont{K.}~\bibnamefont{Huang}},
  \emph{\bibinfo{title}{Statistical Mechanics}} (\bibinfo{publisher}{Wiley},
  \bibinfo{address}{New York}, \bibinfo{year}{1987}).

\bibitem[{\citenamefont{Ellis et~al.}(2004)\citenamefont{Ellis, Touchette, and
  Turkington}}]{touchette2003}
\bibinfo{author}{\bibfnamefont{R.~S.} \bibnamefont{Ellis}},
  \bibinfo{author}{\bibfnamefont{H.}~\bibnamefont{Touchette}},
  \bibnamefont{and}
  \bibinfo{author}{\bibfnamefont{B.}~\bibnamefont{Turkington}},
  \bibinfo{journal}{Physica A} \textbf{\bibinfo{volume}{335}},
  \bibinfo{pages}{518} (\bibinfo{year}{2004}).

\bibitem[{\citenamefont{Tremblay}(1986)}]{tremblay1986}
\bibinfo{author}{\bibfnamefont{A.-M.~S.} \bibnamefont{Tremblay}},
  \bibinfo{journal}{Phys. Lett. A} \textbf{\bibinfo{volume}{116}},
  \bibinfo{pages}{329} (\bibinfo{year}{1986}).

\bibitem[{\citenamefont{Fourcade and Tremblay}(1987)}]{fourcade1987}
\bibinfo{author}{\bibfnamefont{B.}~\bibnamefont{Fourcade}} \bibnamefont{and}
  \bibinfo{author}{\bibfnamefont{A.-M.~S.} \bibnamefont{Tremblay}},
  \bibinfo{journal}{Phys. Rev. A} \textbf{\bibinfo{volume}{36}},
  \bibinfo{pages}{2352} (\bibinfo{year}{1987}).

\bibitem[{\citenamefont{Sz{\'e}pfalusy
  et~al.}(1987)\citenamefont{Sz{\'e}pfalusy, T{\'e}l, Csord{\'a}s, and
  Kovas}}]{szepfalusy1987}
\bibinfo{author}{\bibfnamefont{P.}~\bibnamefont{Sz{\'e}pfalusy}},
  \bibinfo{author}{\bibfnamefont{T.}~\bibnamefont{T{\'e}l}},
  \bibinfo{author}{\bibfnamefont{A.}~\bibnamefont{Csord{\'a}s}},
  \bibnamefont{and} \bibinfo{author}{\bibfnamefont{Z.}~\bibnamefont{Kovas}},
  \bibinfo{journal}{Phys. Rev. A} \textbf{\bibinfo{volume}{36}},
  \bibinfo{pages}{3525} (\bibinfo{year}{1987}).

\bibitem[{\citenamefont{Csord{\'a}s and Sz{\'e}pfalusy}(1989)}]{csordas1989}
\bibinfo{author}{\bibfnamefont{A.}~\bibnamefont{Csord{\'a}s}} \bibnamefont{and}
  \bibinfo{author}{\bibfnamefont{P.}~\bibnamefont{Sz{\'e}pfalusy}},
  \bibinfo{journal}{Phys. Rev. A} \textbf{\bibinfo{volume}{39}},
  \bibinfo{pages}{4767} (\bibinfo{year}{1989}).

\bibitem[{\citenamefont{Jensen}(1988)}]{jensen1988}
\bibinfo{author}{\bibfnamefont{M.~H.} \bibnamefont{Jensen}}, in
  \emph{\bibinfo{booktitle}{Universalities in Condensed Matter}}, edited by
  \bibinfo{editor}{\bibfnamefont{R.}~\bibnamefont{Jullien}},
  \bibinfo{editor}{\bibfnamefont{L.}~\bibnamefont{Peliti}},
  \bibinfo{editor}{\bibfnamefont{R.}~\bibnamefont{Rammal}}, \bibnamefont{and}
  \bibinfo{editor}{\bibfnamefont{N.}~\bibnamefont{Boccara}}
  (\bibinfo{publisher}{Springer}, \bibinfo{address}{Heidelberg},
  \bibinfo{year}{1988}), pp. \bibinfo{pages}{233--235}.

\bibitem[{\citenamefont{Xu et~al.}(2006)\citenamefont{Xu, Ouellette, and
  Bodenschatz}}]{xu2006}
\bibinfo{author}{\bibfnamefont{H.}~\bibnamefont{Xu}},
  \bibinfo{author}{\bibfnamefont{N.}~\bibnamefont{Ouellette}},
  \bibnamefont{and}
  \bibinfo{author}{\bibfnamefont{E.}~\bibnamefont{Bodenschatz}},
  \bibinfo{journal}{Phys. Rev. Lett.} \textbf{\bibinfo{volume}{96}},
  \bibinfo{pages}{114503} (\bibinfo{year}{2006}).

\bibitem[{\citenamefont{Witten and Sander}(1981)}]{witten1981}
\bibinfo{author}{\bibfnamefont{T.~A.} \bibnamefont{Witten}} \bibnamefont{and}
  \bibinfo{author}{\bibfnamefont{L.~M.} \bibnamefont{Sander}},
  \bibinfo{journal}{Phys. Rev. Lett.} \textbf{\bibinfo{volume}{47}},
  \bibinfo{pages}{1400} (\bibinfo{year}{1981}).

\bibitem[{\citenamefont{Halsey et~al.}(1986{\natexlab{b}})\citenamefont{Halsey,
  Meakin, and Procaccia}}]{halsey21986}
\bibinfo{author}{\bibfnamefont{T.~C.} \bibnamefont{Halsey}},
  \bibinfo{author}{\bibfnamefont{P.}~\bibnamefont{Meakin}}, \bibnamefont{and}
  \bibinfo{author}{\bibfnamefont{I.}~\bibnamefont{Procaccia}},
  \bibinfo{journal}{Phys. Rev. Lett.} \textbf{\bibinfo{volume}{56}},
  \bibinfo{pages}{854} (\bibinfo{year}{1986}{\natexlab{b}}).

\bibitem[{\citenamefont{Jensen et~al.}(2002)\citenamefont{Jensen, Levermann,
  Mathiesen, and Procaccia}}]{jensen2002}
\bibinfo{author}{\bibfnamefont{M.~H.} \bibnamefont{Jensen}},
  \bibinfo{author}{\bibfnamefont{A.}~\bibnamefont{Levermann}},
  \bibinfo{author}{\bibfnamefont{J.}~\bibnamefont{Mathiesen}},
  \bibnamefont{and}
  \bibinfo{author}{\bibfnamefont{I.}~\bibnamefont{Procaccia}},
  \bibinfo{journal}{Phys. Rev. E} \textbf{\bibinfo{volume}{65}},
  \bibinfo{pages}{046109} (\bibinfo{year}{2002}).

\bibitem[{\citenamefont{Hata et~al.}(1989)\citenamefont{Hata, Horita, Mori,
  Morita, and Tomita}}]{hata1989}
\bibinfo{author}{\bibfnamefont{H.}~\bibnamefont{Hata}},
  \bibinfo{author}{\bibfnamefont{T.}~\bibnamefont{Horita}},
  \bibinfo{author}{\bibfnamefont{H.}~\bibnamefont{Mori}},
  \bibinfo{author}{\bibfnamefont{T.}~\bibnamefont{Morita}}, \bibnamefont{and}
  \bibinfo{author}{\bibfnamefont{K.}~\bibnamefont{Tomita}},
  \bibinfo{journal}{Prog. Theoret. Phys.} \textbf{\bibinfo{volume}{81}},
  \bibinfo{pages}{11} (\bibinfo{year}{1989}).

\bibitem[{\citenamefont{Mori et~al.}(1989)\citenamefont{Mori, Hata, Horita, and
  Kobayashi}}]{mori1989}
\bibinfo{author}{\bibfnamefont{H.}~\bibnamefont{Mori}},
  \bibinfo{author}{\bibfnamefont{H.}~\bibnamefont{Hata}},
  \bibinfo{author}{\bibfnamefont{T.}~\bibnamefont{Horita}}, \bibnamefont{and}
  \bibinfo{author}{\bibfnamefont{T.}~\bibnamefont{Kobayashi}},
  \bibinfo{journal}{Prog. Theoret. Phys. Suppl.} \textbf{\bibinfo{volume}{99}},
  \bibinfo{pages}{1} (\bibinfo{year}{1989}).

\bibitem[{\citenamefont{Tominaga et~al.}(1990)\citenamefont{Tominaga, Hata,
  Horita, Mori, and Tomita}}]{tominaga1990}
\bibinfo{author}{\bibfnamefont{H.}~\bibnamefont{Tominaga}},
  \bibinfo{author}{\bibfnamefont{H.}~\bibnamefont{Hata}},
  \bibinfo{author}{\bibfnamefont{T.}~\bibnamefont{Horita}},
  \bibinfo{author}{\bibfnamefont{H.}~\bibnamefont{Mori}}, \bibnamefont{and}
  \bibinfo{author}{\bibfnamefont{K.}~\bibnamefont{Tomita}},
  \bibinfo{journal}{Prog. Theoret. Phys.} \textbf{\bibinfo{volume}{84}},
  \bibinfo{pages}{18} (\bibinfo{year}{1990}).

\bibitem[{\citenamefont{Kastner}(2000)}]{kastner2000}
\bibinfo{author}{\bibfnamefont{M.}~\bibnamefont{Kastner}}, Ph.D. thesis,
  \bibinfo{school}{Universit\"at Erlangen-N\"urnberg} (\bibinfo{year}{2000}).

\bibitem[{\citenamefont{Pleimling et~al.}(2004)\citenamefont{Pleimling,
  Behringer, and H\"uller}}]{pleimling2004}
\bibinfo{author}{\bibfnamefont{M.}~\bibnamefont{Pleimling}},
  \bibinfo{author}{\bibfnamefont{H.}~\bibnamefont{Behringer}},
  \bibnamefont{and} \bibinfo{author}{\bibfnamefont{A.}~\bibnamefont{H\"uller}},
  \bibinfo{journal}{Phys. Lett. A} \textbf{\bibinfo{volume}{328}},
  \bibinfo{pages}{432} (\bibinfo{year}{2004}).

\bibitem[{\citenamefont{Behringer et~al.}(2005)\citenamefont{Behringer,
  Pleimling, and H\"uller}}]{behringer2005}
\bibinfo{author}{\bibfnamefont{H.}~\bibnamefont{Behringer}},
  \bibinfo{author}{\bibfnamefont{M.}~\bibnamefont{Pleimling}},
  \bibnamefont{and} \bibinfo{author}{\bibfnamefont{A.}~\bibnamefont{H\"uller}},
  \bibinfo{journal}{J. Phys. A} \textbf{\bibinfo{volume}{38}},
  \bibinfo{pages}{973} (\bibinfo{year}{2005}).

\bibitem[{\citenamefont{Sano et~al.}(1986)\citenamefont{Sano, Sato, and
  Sawada}}]{sano1986}
\bibinfo{author}{\bibfnamefont{M.}~\bibnamefont{Sano}},
  \bibinfo{author}{\bibfnamefont{S.}~\bibnamefont{Sato}}, \bibnamefont{and}
  \bibinfo{author}{\bibfnamefont{Y.}~\bibnamefont{Sawada}},
  \bibinfo{journal}{Prog. Theoret. Phys.} \textbf{\bibinfo{volume}{76}},
  \bibinfo{pages}{945} (\bibinfo{year}{1986}).

\bibitem[{\citenamefont{Hata et~al.}(1988)\citenamefont{Hata, Horita, Mori,
  Morita, and Tomita}}]{hata1988}
\bibinfo{author}{\bibfnamefont{H.}~\bibnamefont{Hata}},
  \bibinfo{author}{\bibfnamefont{T.}~\bibnamefont{Horita}},
  \bibinfo{author}{\bibfnamefont{H.}~\bibnamefont{Mori}},
  \bibinfo{author}{\bibfnamefont{T.}~\bibnamefont{Morita}}, \bibnamefont{and}
  \bibinfo{author}{\bibfnamefont{K.}~\bibnamefont{Tomita}},
  \bibinfo{journal}{Prog. Theoret. Phys.} \textbf{\bibinfo{volume}{80}},
  \bibinfo{pages}{809} (\bibinfo{year}{1988}).

\bibitem[{\citenamefont{Horita et~al.}(1988)\citenamefont{Horita, Hata, Mori,
  Morita, Tomita, Kuroki, and Okamoto}}]{horita1988}
\bibinfo{author}{\bibfnamefont{T.}~\bibnamefont{Horita}},
  \bibinfo{author}{\bibfnamefont{H.}~\bibnamefont{Hata}},
  \bibinfo{author}{\bibfnamefont{H.}~\bibnamefont{Mori}},
  \bibinfo{author}{\bibfnamefont{T.}~\bibnamefont{Morita}},
  \bibinfo{author}{\bibfnamefont{K.}~\bibnamefont{Tomita}},
  \bibinfo{author}{\bibfnamefont{S.}~\bibnamefont{Kuroki}}, \bibnamefont{and}
  \bibinfo{author}{\bibfnamefont{H.}~\bibnamefont{Okamoto}},
  \bibinfo{journal}{Prog. Theoret. Phys.} \textbf{\bibinfo{volume}{80}},
  \bibinfo{pages}{793} (\bibinfo{year}{1988}).

\bibitem[{\citenamefont{Tomita et~al.}(1988)\citenamefont{Tomita, Hata, Horita,
  Mori, and Morita}}]{tomita1988}
\bibinfo{author}{\bibfnamefont{K.}~\bibnamefont{Tomita}},
  \bibinfo{author}{\bibfnamefont{H.}~\bibnamefont{Hata}},
  \bibinfo{author}{\bibfnamefont{T.}~\bibnamefont{Horita}},
  \bibinfo{author}{\bibfnamefont{H.}~\bibnamefont{Mori}}, \bibnamefont{and}
  \bibinfo{author}{\bibfnamefont{T.}~\bibnamefont{Morita}},
  \bibinfo{journal}{Prog. Theoret. Phys.} \textbf{\bibinfo{volume}{80}},
  \bibinfo{pages}{953} (\bibinfo{year}{1988}).

\bibitem[{\citenamefont{Tomita et~al.}(1989)\citenamefont{Tomita, Hata, Horita,
  Mori, Morita, Okamoto, and Tominaga}}]{tomita1989}
\bibinfo{author}{\bibfnamefont{K.}~\bibnamefont{Tomita}},
  \bibinfo{author}{\bibfnamefont{H.}~\bibnamefont{Hata}},
  \bibinfo{author}{\bibfnamefont{T.}~\bibnamefont{Horita}},
  \bibinfo{author}{\bibfnamefont{H.}~\bibnamefont{Mori}},
  \bibinfo{author}{\bibfnamefont{T.}~\bibnamefont{Morita}},
  \bibinfo{author}{\bibfnamefont{H.}~\bibnamefont{Okamoto}}, \bibnamefont{and}
  \bibinfo{author}{\bibfnamefont{H.}~\bibnamefont{Tominaga}},
  \bibinfo{journal}{Prog. Theoret. Phys.} \textbf{\bibinfo{volume}{81}},
  \bibinfo{pages}{1124} (\bibinfo{year}{1989}).

\bibitem[{\citenamefont{Yoshida and Miyazaki}(1989)}]{yoshida1989}
\bibinfo{author}{\bibfnamefont{T.}~\bibnamefont{Yoshida}} \bibnamefont{and}
  \bibinfo{author}{\bibfnamefont{S.}~\bibnamefont{Miyazaki}},
  \bibinfo{journal}{Prog. Theoret. Phys. Suppl.} \textbf{\bibinfo{volume}{99}},
  \bibinfo{pages}{64} (\bibinfo{year}{1989}).

\bibitem[{\citenamefont{Costeniuc et~al.}(2006)\citenamefont{Costeniuc, Ellis,
  Touchette, and Turkington}}]{costeniuc2006}
\bibinfo{author}{\bibfnamefont{M.}~\bibnamefont{Costeniuc}},
  \bibinfo{author}{\bibfnamefont{R.~S.} \bibnamefont{Ellis}},
  \bibinfo{author}{\bibfnamefont{H.}~\bibnamefont{Touchette}},
  \bibnamefont{and}
  \bibinfo{author}{\bibfnamefont{B.}~\bibnamefont{Turkington}},
  \bibinfo{journal}{Phys. Rev. E} \textbf{\bibinfo{volume}{73}},
  \bibinfo{pages}{026105} (\bibinfo{year}{2006}).

\bibitem[{\citenamefont{Ellis et~al.}(2000)\citenamefont{Ellis, Haven, and
  Turkington}}]{ellis2000}
\bibinfo{author}{\bibfnamefont{R.~S.} \bibnamefont{Ellis}},
  \bibinfo{author}{\bibfnamefont{K.}~\bibnamefont{Haven}}, \bibnamefont{and}
  \bibinfo{author}{\bibfnamefont{B.}~\bibnamefont{Turkington}},
  \bibinfo{journal}{J. Stat. Phys.} \textbf{\bibinfo{volume}{101}},
  \bibinfo{pages}{999} (\bibinfo{year}{2000}).

\bibitem[{\citenamefont{Touchette et~al.}(2004)\citenamefont{Touchette, Ellis,
  and Turkington}}]{touchette2004}
\bibinfo{author}{\bibfnamefont{H.}~\bibnamefont{Touchette}},
  \bibinfo{author}{\bibfnamefont{R.~S.} \bibnamefont{Ellis}}, \bibnamefont{and}
  \bibinfo{author}{\bibfnamefont{B.}~\bibnamefont{Turkington}},
  \bibinfo{journal}{Physica A} \textbf{\bibinfo{volume}{340}},
  \bibinfo{pages}{138} (\bibinfo{year}{2004}).

\bibitem[{\citenamefont{Dinwoodie and Zabell}(1992)}]{dinwoodie1992}
\bibinfo{author}{\bibfnamefont{I.~H.} \bibnamefont{Dinwoodie}}
  \bibnamefont{and} \bibinfo{author}{\bibfnamefont{S.~L.}
  \bibnamefont{Zabell}}, \bibinfo{journal}{Ann. Prob.}
  \textbf{\bibinfo{volume}{20}}, \bibinfo{pages}{1147} (\bibinfo{year}{1992}).

\bibitem[{\citenamefont{Ioffe}(1993)}]{ioffe1993}
\bibinfo{author}{\bibfnamefont{D.}~\bibnamefont{Ioffe}},
  \bibinfo{journal}{Stat. Prob. Lett.} \textbf{\bibinfo{volume}{18}},
  \bibinfo{pages}{297} (\bibinfo{year}{1993}).

\bibitem[{\citenamefont{Ellis}(1995)}]{ellis1995}
\bibinfo{author}{\bibfnamefont{R.~S.} \bibnamefont{Ellis}},
  \bibinfo{journal}{Scand. Actuarial J.} \textbf{\bibinfo{volume}{1}},
  \bibinfo{pages}{97} (\bibinfo{year}{1995}).

\bibitem[{\citenamefont{Touchette}(2003)}]{touchette22003}
\bibinfo{author}{\bibfnamefont{H.}~\bibnamefont{Touchette}}, Ph.D. thesis,
  \bibinfo{school}{McGill University} (\bibinfo{year}{2003}).

\bibitem[{\citenamefont{Plachky and Steinebach}(1975)}]{plachky1975}
\bibinfo{author}{\bibfnamefont{D.}~\bibnamefont{Plachky}} \bibnamefont{and}
  \bibinfo{author}{\bibfnamefont{J.}~\bibnamefont{Steinebach}},
  \bibinfo{journal}{Per. Math. Hung.} \textbf{\bibinfo{volume}{6}},
  \bibinfo{pages}{343} (\bibinfo{year}{1975}).

\bibitem[{\citenamefont{Lanford}(1973)}]{lanford1973}
\bibinfo{author}{\bibfnamefont{O.~E.} \bibnamefont{Lanford}}, in
  \emph{\bibinfo{booktitle}{Statistical Mechanics and Mathematical Problems}},
  edited by \bibinfo{editor}{\bibfnamefont{A.}~\bibnamefont{Lenard}}
  (\bibinfo{publisher}{Springer}, \bibinfo{address}{Berlin},
  \bibinfo{year}{1973}), vol.~\bibinfo{volume}{20} of
  \emph{\bibinfo{series}{Lecture Notes in Physics}}, pp.
  \bibinfo{pages}{1--113}.

\bibitem[{\citenamefont{Zohar}(1999)}]{zohar1999}
\bibinfo{author}{\bibfnamefont{G.}~\bibnamefont{Zohar}},
  \bibinfo{journal}{Stoc. Proc. Appl.} \textbf{\bibinfo{volume}{79}},
  \bibinfo{pages}{229} (\bibinfo{year}{1999}).

\bibitem[{\citenamefont{Veneziano}(2002)}]{veneziano2002}
\bibinfo{author}{\bibfnamefont{D.}~\bibnamefont{Veneziano}},
  \bibinfo{journal}{Fractals} \textbf{\bibinfo{volume}{10}},
  \bibinfo{pages}{117} (\bibinfo{year}{2002}).

\bibitem[{\citenamefont{Oono}(1989)}]{oono1989}
\bibinfo{author}{\bibfnamefont{Y.}~\bibnamefont{Oono}}, \bibinfo{journal}{Prog.
  Theoret. Phys. Suppl.} \textbf{\bibinfo{volume}{99}}, \bibinfo{pages}{165}
  (\bibinfo{year}{1989}).

\bibitem[{\citenamefont{Ruelle}(2004)}]{ruelle2004}
\bibinfo{author}{\bibfnamefont{D.}~\bibnamefont{Ruelle}},
  \emph{\bibinfo{title}{Thermodynamic Formalism}}
  (\bibinfo{publisher}{Cambridge University Press},
  \bibinfo{address}{Cambridge}, \bibinfo{year}{2004}), \bibinfo{edition}{2nd}
  ed.

\bibitem[{\citenamefont{Imparato and Peliti}(2005)}]{imparato2005}
\bibinfo{author}{\bibfnamefont{A.}~\bibnamefont{Imparato}} \bibnamefont{and}
  \bibinfo{author}{\bibfnamefont{L.}~\bibnamefont{Peliti}},
  \bibinfo{journal}{Phys. Rev. E} \textbf{\bibinfo{volume}{72}},
  \bibinfo{pages}{046114} (\bibinfo{year}{2005}).

\bibitem[{\citenamefont{Testud}(2005)}]{testud2005}
\bibinfo{author}{\bibfnamefont{B.}~\bibnamefont{Testud}}, \bibinfo{journal}{C.
  R. Acad. Sci. Paris, Ser. I} \textbf{\bibinfo{volume}{340}},
  \bibinfo{pages}{653} (\bibinfo{year}{2005}).

\bibitem[{\citenamefont{Testud}(2006)}]{testud2006}
\bibinfo{author}{\bibfnamefont{B.}~\bibnamefont{Testud}},
  \bibinfo{journal}{Nonlinearity} \textbf{\bibinfo{volume}{19}},
  \bibinfo{pages}{1201} (\bibinfo{year}{2006}).

\bibitem[{\citenamefont{Riedi}(1995)}]{riedi1995}
\bibinfo{author}{\bibfnamefont{R.}~\bibnamefont{Riedi}}, \bibinfo{journal}{J.
  Math. Anal. Appl.} \textbf{\bibinfo{volume}{189}}, \bibinfo{pages}{462}
  (\bibinfo{year}{1995}).

\end{thebibliography}

\end{document}